\documentclass[useAMS,usenatbib]{mn2e}
\usepackage{amssymb, amsmath, epsfig}
\newcommand{\Mpc}{{\rm Mpc}}

\newcommand{\cmeter}{{\rm cm}}
\newcommand{\Msun}{{M_{\odot}}}
\def\apj{ApJ}
\def\apjl{ApJL}
\def\apjs{ApJS}
\def\aj{AJ}
\def\mnras{MNRAS}

\def\pasj{PASJ}

\def\aap{{\em A.\&A}}

\begin{document}

\title{Probing the Neutral Fraction of the IGM with GRBs during the Epoch of Reionisation}

\author[M. McQuinn et al.]{Matthew McQuinn$^1$\thanks{mmcquinn@cfa.harvard.edu}, Adam Lidz$^{1}$,
Matias Zaldarriaga$^{1,2}$, Lars Hernquist$^{1}$,\newauthor
Suvendra Dutta$^{1}$ \\
$^{1}$ Harvard-Smithsonian Center
for Astrophysics, 60 Garden St., Cambridge, MA 02138\\
$^2$  Jefferson Laboratory of Physics, Harvard University,
Cambridge, MA 02138\\
}

\pubyear{2006} \volume{000} \pagerange{1}

\maketitle\label{firstpage}

\begin{abstract}

We show that near-infrared observations of the red side of the
Ly$\alpha$ line from a single gamma ray burst (GRB) afterglow cannot
be used to constrain the global neutral fraction of the intergalactic
medium (IGM), $\bar{x}_H$, at the GRB's redshift to better than
$\delta \bar{x}_H \sim 0.3$.  Some GRB sight-lines will encounter more
neutral hydrogen than others at fixed $\bar{x}_H$ owing to the
patchiness of reionisation.  GRBs during the epoch of reionisation
will often bear no discernible signature of a neutral IGM in their
afterglow spectra.  We discuss the constraints on $\bar{x}_H$ from the
$z = 6.3$ burst, GRB050904, and quantify the probability of detecting
a neutral IGM using future spectroscopic observations of
high-redshift, near-infrared GRB afterglows.  Assuming an observation
with signal-to-noise similar to the Subaru FOCAS spectrum of GRB050904
and that the column density distribution of damped Ly$\alpha$
absorbers is the same as measured at lower redshifts, a GRB from an
epoch when $\bar{x}_H = 0.5$ can be used to detect a partly neutral
IGM at $97\%$ confidence level $\approx 10$\% of the time (and, for an
observation with three times the sensitivity, $\approx 30$\% of the
time).

\end{abstract}


\section{Introduction}

Tomorrow, a gamma ray burst (GRB) may be observed that originates from
the death of one of the first stars, during the epoch of reionisation.
Despite the great distance to this burst, it will be the brightest
gamma ray source on the sky for several tens of seconds, one of the
brightest cosmological X-ray sources for hours, and its afterglow will
be observable for weeks in the near-infrared (and, for the first few
hours, brighter than any $z \sim 6$ QSO).  Much of the optical and
near-infrared light will be obscured by the Ly$\alpha$ forest, and
this obscuration will enable the strongest constraint to date on the
neutral hydrogen fraction of the intergalactic medium (IGM) at the
burst's redshift.

In fact, such an occurrence may already have been realised.
\citet{haislip06}, \citet{kawai05}, \citet{tagliaferri05}, and
\citet{totani06} observed and analysed the optical/near-infrared
afterglow of GRB050904, identified to be at $z =6.3$ -- possibly
during the reionisation epoch and the GRB with the highest confirmed
redshift.  \citet{totani06} derived the constraint on the global
neutral fraction $\bar{x}_{H} < 0.6$ at $z = 6.3$.  In this paper, we
discuss the assumptions that went into their analysis, and we
investigate how realistic modelling of reionisation can affect
constraints on $\bar{x}_H$ from GRB050904 and from future $z>6$ GRBs.

The Swift satellite has greatly increased the sample of GRBs with
known redshifts in the last two years \citep{gehrels04}. Future
missions such as EXIST \citep{grindlay06} and JWST \citep{gardner06}
will further enhance our ability to detect high-redshift GRBs and will
enable more detailed follow-up studies of their near-infrared
afterglows.  Interestingly, approximately one-half of Swift bursts are
``dark bursts'' -- bursts that have detected X-ray afterglows, but
that have no measurable optical emission (e.g.,
\citealt{filliatre06}).  While it is probable that most dark bursts
originate from low-redshift, dust-rich galaxies, a fraction of dark
bursts may originate from $z>6$ and are ``dark'' because Ly$\alpha$
absorption from the high-redshift IGM absorbs the optical emission
(e.g., \citealt{malesani05}).

In addition to their extreme luminosity, there are several other
advantages to studying reionisation with GRBs compared to other probes
of this epoch.  First, the afterglows of high-redshift GRBs are
observed at earlier (brighter) times in the source frame than those at
lower redshifts, so the dimming owing to increased luminosity distance
is nearly cancelled, and the observed flux is almost independent of
redshift \citep{lamb01, ciardi99}.  Second, unlike the spectra of
galaxies and quasars, the intrinsic afterglow spectrum of a GRB is a
featureless power-law at the relevant wavelengths, allowing a more
precise measurement of absorption owing to a neutral IGM
\citep{barkana04b}.  Finally, since the theoretical expectation is
that most of the star formation at $z \gtrsim 6$ occurs in halos with
$m \sim 10^9 \; \Msun$ and because observations at $z \gtrsim 6$
currently probe only the most massive galaxies and QSOs ($m \gtrsim
10^{11} ~\Msun$), high-redshift GRB host galaxies should be less
massive than galaxies selected in another manner.  Consequently, GRB
host galaxies will sit in smaller HII regions during reionisation (on
average) than galaxies selected by different means.  Therefore, GRBs
will suffer a larger Ly$\alpha$ IGM absorption feature.

In this work, we do not concentrate on wavelengths blueward of
source-frame Ly$\alpha$ (in the Ly$\alpha$ forest) to derive
constraints from GRBs.  Any blueward flux indicates the presence of
ionised gas at the redshift of the transmission.  However, at high
redshifts there is little or no flux in the Ly$\alpha$ forest, even in
ionised regions, owing to the increase in density with increasing
redshift, the decrease in the size and in the number of voids, and the
decrease in the amplitude of the ionising background (e.g.,
\citealt{becker06} and \citealt{lidz07}).  As a result, it is difficult
to distinguish a partly ionised IGM from a fully ionised one with the
$z > 6$ forest \citep{becker06, lidz07}.  Future observations of the
$z >6$ Ly$\alpha$ forest from additional QSOs and GRBs will aid
reionisation studies, but is unclear whether such studies will ever
provide definitive evidence for neutral pockets in the IGM.
In contrast, the shape of the line profile redward of Ly$\alpha$
\emph{is} sensitive to a substantially neutral IGM and, therefore, can be
used to unambiguously detect reionisation \citep{miralda98}.

Little is known about the rate of GRBs at $z > 6$.  We assume that the
  rate of long GRBs traces the massive star formation rate (SFR) for
  most calculations in this work.\footnote{We do not consider the
  other class of GRBs, the ``short'' GRBs, in this study.  These
  bursts, while still cosmological, are more local than long GRBs and
  typically do not have a detected afterglow.}  The
  assumption that the GRB rate traces the massive SFR is supported by
  observations of lower redshift GRB host galaxies \citep{bloom02,
  djorgovski01}.  However, \citet{kistler07} found that the GRB rate
  is four times higher at $z \approx 4$ than if the GRB rate exactly
  traces the SFR.  Other properties of a galaxy apart from its massive
  SFR might be correlated with its rate of GRBs.  For example,
  \citet{stanek06} found that $z < 0.25$ GRBs -- GRBs that are
  typically under-luminous -- are preferentially in metal-poor
  galaxies.

Making the assumption that the GRB rate traces the observed SFR,
  \citet{salvaterra08} predicted that SWIFT will be triggered by $1-4$
  bursts a year above $z = 6$\footnote{This rate is larger than the
  SWIFT rate of $1$ identified $z >6$ bursts in three years, but it is
  possible that these numbers can be reconciled in light of these dark
  bursts.} and that the EXIST mission would observe $10-60$ bursts a
  year.  Other studies have predicted even larger rates
  \citep{bromm02, daigne06}.

At $z > 6$, POPIII stars with average masses of $\sim 100~\Msun$ may
exist.  It is unclear whether the death of a POPIII star can result in
a GRB.  \citet{fryer01} identified a mechanism that might produce GRBs
from POPIII stars.  However, once the interstellar metallicity reaches
a critical value of $\sim 10^{-3.5}$ solar in a high-redshift galaxy,
POPIII star formation quenches and the normal mode of POPII star
formation begins (e.g., \citealt{mackey03, yoshida04}), and this mode
{\it is} known to produce GRBs.  Most if not all of reionisation
likely owes to photons from POPII-like stars (e.g.,
\citealt{sokasian04, trac06}).

In Section \ref{redwing}, we discuss the absorption profiles of a
neutral IGM as well as of a damped Ly$\alpha$ absorber (DLA), and, in
Section \ref{reionisation}, we discuss our simulations of reionisation
and their implications for the amount of IGM absorption in a GRB
afterglow spectrum.  Section \ref{fits} quantifies the detectability
of a neutral IGM from GRB afterglow spectra, and Section
\ref{GRB050904} describes the constraints on $\bar{x}_H$ from the $z =
6.3$ burst, GRB050904.  For our calculations, we adopt a cosmology
with $\Omega_m = 0.27$, $\Omega_\Lambda = 0.73$, $\Omega_b = 0.46$,
$\sigma_8 = 0.8$, $n = 1$, and $h = 0.7$, which is consistent with the
most recent cosmic microwave background and large scale structure data
\citep{spergel06}.  We express all distances in comoving units unless
otherwise noted.

When this project was nearing completion, we learned of a similar
effort by \citet{mesinger07b} and refer the reader there for a
complementary discussion.

\section{The Red Damping Wing}
\label{redwing}

In the standard picture, GRB afterglows result from shells of
relativistic matter colliding with the interstellar medium, shocking,
and radiating via synchrotron emission \citep{sari98}.  The observed
flux of a GRB just redward of GRB-frame Ly$\alpha$ is given roughly
by ${\cal F} = A \; (\nu/\nu_{\alpha})^{-\beta} \;
\exp[-\tau_{\alpha}(\nu)]$, where $A$ is a time-dependent amplitude,
$\beta$ is a time-dependent power-law index, $\nu_\alpha$ is the
frequency at the Ly$\alpha$ line centre, and
\begin{eqnarray}
\tau_{\alpha}(\nu) & \approx & \int_0^{z_{g}} \; \frac{dz}{1+z} \, \frac{c}{H(z)} \, n_H(z) \; \sigma_{\alpha}\left( \nu_z \, (1+\frac{v(z)}{c}) \right) \nonumber \\
 &+&  N_{\rm HI} \; \sigma_{\alpha} \left(\nu_{z_{\rm DLA}} \,(1+\frac{v(z_{\rm DLA})}{c})\right).
 \label{eqn:tau}
\end{eqnarray}
 Here, $\nu_z \equiv \nu \;(1 +z)$, $z_g$ [$z_{\rm DLA}$] is the
redshift of the GRB host galaxy [DLA], $\sigma_{\alpha}(\nu)$ is the
Lorentzian-like Ly$\alpha$ naturally broadened scattering cross
section at frequency $\nu$, and $v(z)$ is the line-of-sight peculiar
velocity of the gas.  The first term on the right-hand side of
Equation (\ref{eqn:tau}) owes to IGM absorption and the second owes to
absorption by a DLA with column density $N_{\rm HI}$.  The extended
absorption profile redward of Ly$\alpha$ shaped by the factor
$\exp[-\tau_{\alpha}(\nu)]$ is commonly referred to as the ``damping
wing''.


Let us first ignore absorption owing to a DLA and the effects of
peculiar velocities.  For an isolated bubble of comoving size $R_b$ in
a homogeneous medium with neutral fraction $\bar{x}_H$, Equation
(\ref{eqn:tau}) simplifies to the expression (e.g., \citealt{loeb99})

\begin{eqnarray}
\tau_{\alpha}(\nu)  \approx 900 \;{\rm km \, s^{-1}} &\times& \bar{x}_H \,
\left(\frac{1+z_g}{8}\right)^{3/2} \nonumber \\
& \times& \left(\frac{H(z_g) \, R_b}{(1+z_g)} - c \,\frac{\nu_{z}- \nu_\alpha}{\nu_\alpha} \right)^{-1},
\label{eqn:tau2}
\end{eqnarray}
or $\tau(\nu_{\alpha}/(1+z_g)) \approx \bar{x}_H$ for a $1$ proper Mpc
bubble, noting that $H(z_g = 7) \approx 900$ km s$^{-1}$ proper
Mpc$^{-1}$.\footnote{To derive Equation (\ref{eqn:tau2}), approximate
$\sigma_\alpha(\nu)$ with a Lorentzian profile and ignore the redshift
dependence that appears in the numerator of the integrand in Equation
(\ref{eqn:tau}).  We have checked that these approximations are
accurate to better than $10\%$ for relevant $R_b$.}  The width of the
damping wing feature from a neutral IGM (defined here as the the
wavelength for which $\exp[-\tau_{\alpha}(\nu)] < 0.9$) is $\approx 25
\;(1 +z_g) \; \AA$.  The width of this feature is typically broader
than the feature due to a DLA, which falls off as ${\Delta
\lambda_{\alpha}}^{-2}$ rather than the ${\Delta
\lambda_{\alpha}}^{-1}$ scaling for IGM absorption, where ${\Delta
\lambda_{\alpha}}$ is the difference between a given wavelength and
that of Ly$\alpha$ in the GRB-frame \citep{miralda98}.

\begin{figure}
{\epsfig{file=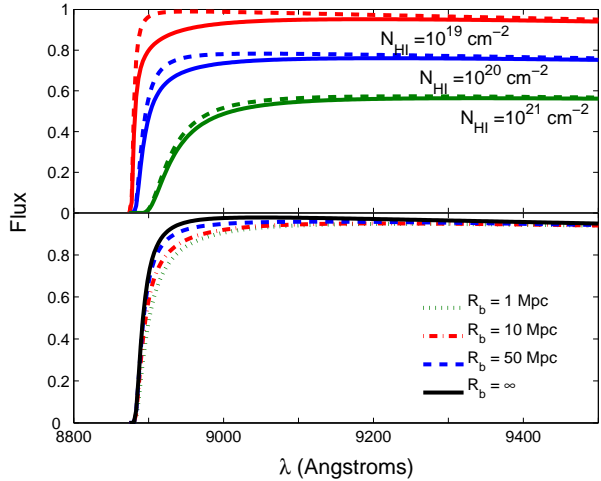, width=9cm}}
\caption{Illustration of the GRB afterglow flux near the GRB-frame
Ly${\alpha}$ wavelength.  The top panel depicts the afterglow flux for
both an ionised universe (dashed curves) and with $\bar{x}_H = 0.5$
and $R_b = 10$ Mpc (solid curves).  The bottom panel illustrates the
effect of bubble size on the afterglow flux, assuming $\bar{x}_H =
0.5$ and $N_{\rm HI} = 10^{20} ~ \cmeter^{-2}$.  The normalisation of
the flux in these panels is arbitrary, and it has been adjusted in the
top panel to separate the sets of curves. These curves are calculated
with $\beta = 1.25$ and $z_g = 6.3$, and $\lambda$ is the observed
wavelength. }
\label{fig:params}
\end{figure}

Figure \ref{fig:params} illustrates the effect of the column density
of the DLA (top panel) and the size of the HII region (bottom panel)
on the absorption feature.  Here we adopt the same simplistic
parameterisation as in Equation (\ref{eqn:tau2}): an HII region of
size $R_b$ surrounded by a homogeneously neutral IGM with neutral
fraction $\bar{x}_H$.  The dashed lines in the top panel represent an
IGM with $\bar{x}_H = 0$, and the solid lines represent the case with
$\bar{x}_H = 0.5$ and $R_b = 10$ Mpc.  IGM absorption produces a wider
feature than that from a DLA.  However, it becomes hard to visually
discern absorption owing to a neutral IGM for $N_{\rm HI} \gtrsim
10^{21}~\cmeter^{-2}$.  This illustration suggests that it is
difficult to measure $\bar{x}_H$ from GRB050904, which has $N_{\rm HI}
\approx 10^{21.6}~\cmeter^{-2}$ \citep{totani06}.  We return to this
point in Section \ref{GRB050904}.

DLAs with $N_{\rm HI} > 10^{19} \; {\rm cm}^{-2}$ are associated with
all but one GRB for which $N_{\rm HI}$ has been measured
\citep{chen07}.  The cumulative distribution of $N_{\rm HI}$ for the
current sample of $\approx 30$ GRB DLAs scales as $N_{\rm HI}^{0.3}$
between $10^{18}$ and $10^{21.5} \;{\rm cm}^{-2}$, and about half of all
GRBs for which $N_{\rm HI}$ has been measured have $N_{\rm HI} >
10^{21.5} \; {\rm cm}^{-2}$ \citep{chen07}.  However, it is not clear
how $N_{\rm HI}$ should scale with redshift.  It is plausible that,
since the average galaxy becomes less massive with redshift, $z >6$ galaxies
should, on average, have weaker DLAs.  For galaxies to reionise the
Universe, the escape fraction of ionising photons must be appreciable,
implying that sight-lines with $N_{\rm HI} \lesssim 10^{18}\; {\rm
cm}^{-2}$ must exist.\footnote{An optical depth of unity at the HI
Lyman-limit requires $N_{\rm HI} = 2\times 10^{17}$ cm$^{-2}$.}

The bottom panel in Figure \ref{fig:params} depicts the effect of
bubble size on the GRB afterglow spectrum.  The IGM absorption for
GRBs in bubbles with sizes between $1$-$10 ~\Mpc$ is comparable.
However, if the GRB sits in a large bubble with $R_b \sim 50 ~\Mpc$,
the absorption is strongly diminished.  When $\bar{x}_H = 0.2$,
roughly half of the skewers from GRBs in our simulations of
reionisation sit in HII regions that are larger than $50$ Mpc (Section
\ref{reionisation}).

\section{Effect of Patchy Reionisation}
\label{reionisation}

\begin{figure}
\begin{center}
{\epsfig{file=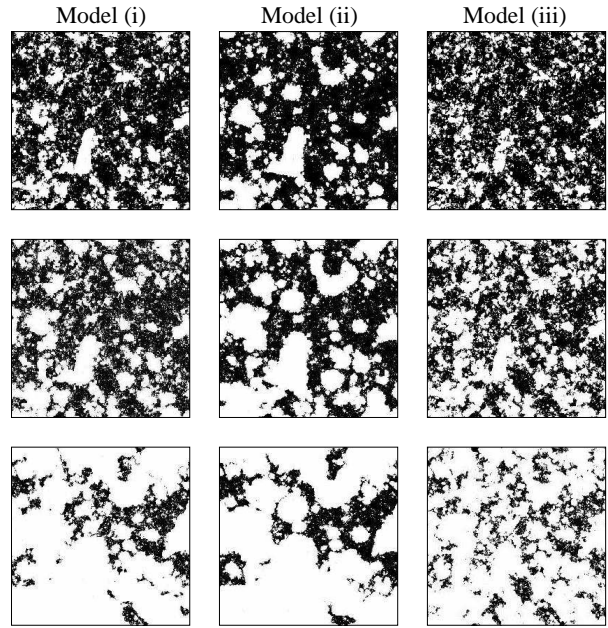, width=8.0cm}}
\end{center}
\caption{Slices through the middle of the simulation box for the three
models discussed in the text, each with width $186 \; \Mpc$ and with
$6.9< z < 8.5$.  The white regions are fully ionised and the black are
fully neutral.  The top row uses snapshots with volume-weighted
neutral fractions of $\bar{x}_{H} = 0.7$, the middle uses those
with $\bar{x}_{H} = 0.5$, and the bottom uses those with
$\bar{x}_{H} = 0.2$.
\label{fig:slice}}
\end{figure}

 While the GRB itself does not ionise the IGM around it, earlier star
formation from its host and neighbouring galaxies can grow a large HII
region.  To model reionisation around a GRB, we employ three radiative
transfer simulations that are each $186 \; \Mpc$ on a side.  These
simulations are post-processed on a $1024^3$ particle N-body field.
Unresolved halos are included with extended Press-Schechter merger
trees.  These three simulations are described in detail in
\citet{mcquinn07} [and the methodology is described in
\citet{mcquinn06b}], and they are meant to span the range of plausible
morphologies for reionisation by galaxies.  Here is a brief
description of the models for the sources and sinks of ionising
photons used in each of the three simulations:
\begin{description}
\item[Model (i):] All halos above the mass at which the gas can cool atomically
($m_{\rm cool}$) contribute ionising photons at a rate that is
proportional to their mass $m$.  The scaling $\dot{N}_{\rm
ion} \sim m$ assumes that the massive SFR is proportional to
the amount of gas within a galaxy.
\item[Model (ii):] Halos more massive than $m_{\rm cool}$ contribute to the
production of ionising photons, with the ionising luminosity of the
sources scaling as halo mass to the $5/3$ power.  This scaling is
chosen to match the relationship between star formation efficiency and
galaxy mass that is observed in low-redshift dwarf galaxies
\citep{kauffmann03} as well as the SFR found in
theoretical studies that include supernova feedback \citep{dekel03,
hernquist03, springel03}.
\item[Model (iii):] Absorption by minihalos shapes the morphology of
reionisation and $\dot{N}_{\rm ion} \sim m$ for $m > m_{\rm cool}$ .
All minihalos with $m > 10^5~M_{\odot}$ absorb incident ionising
photons out to their virial radii until they are photo-evaporated.  We
use the fitting formula for the evaporation timescale given in
\citet{iliev}.  This timescale is roughly the sound-crossing time of
a halo, or $t_{\rm ev} = 100 \; (m/10^7  \Msun) \; {\rm Myr}$ \citep{shapiro}.
\end{description}

The normalisation of the function $\dot{N}_{\rm ion}(m)$ in the three
simulations is chosen such that reionisation is completed by $z
\approx 7$.  Given the uncertainties in $f_{\rm esc}$ -- the fraction
of ionising photons that escape and ionise the IGM -- and in the SFR in
high-redshift galaxies, there is a large range of possible
normalisations.  Fortunately, the morphology of reionisation when
comparing at fixed $\bar{x}_H$ depends only weakly on the
normalisation of $\dot{N}_{\rm ion}$, as shown in \citet{mcquinn06b}.
\citet{mcquinn06b} and \citet{mcquinn07} also showed that other
effects such as thermal feedback on sub-Jeans mass galaxies, source
duty cycle, and the number of recombinations have negligible impact on
the morphology for reasonable models.  In fact, studies have
demonstrated that the morphology of reionisation is shaped principally
by the clustering of the ionising sources \citep{furlanetto04a, fsh04,
furl-models, zahn06, mcquinn06b}.

Figure \ref{fig:slice} displays slices through simulations 
adopting reionisation
models (i), (ii), and (iii).  The white regions are ionised and the
black are neutral.  Model (ii) results in the largest HII regions
because it has the most biased sources, whereas model (iii) produces
the smallest bubbles, with the maximum bubble radius restricted to be
roughly the mean free path for ionising photons to intersect a
minihalo.

\begin{figure}
\rotatebox{-90}{\epsfig{file=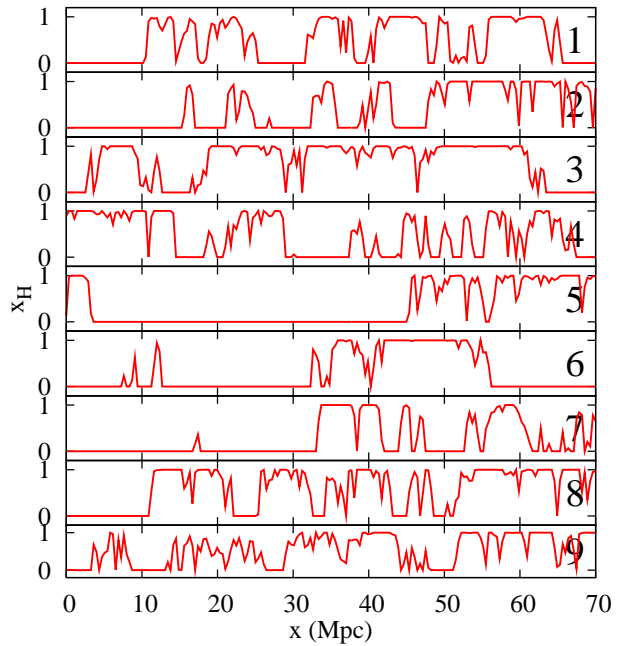, height = 12.3cm}}
\caption{Nine randomly selected $70$ Mpc skewers originating at a GRB
in the simulation volume.  These sight-lines are chosen from a
snapshot from the simulation of model (i) that has $\bar{x}_{H} = 0.5$
and is from $z= 7.3$.
\label{fig:skewer}}
\end{figure}

\begin{figure}
\rotatebox{-90}{\epsfig{file=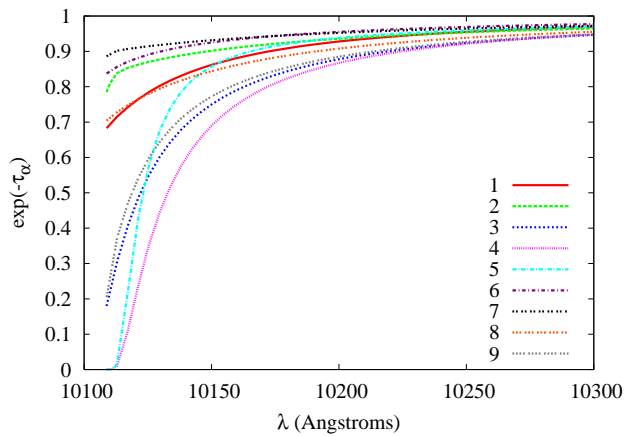, height=8.4cm}}
\caption{Transmission for the nine skewers displayed in
Figure \ref{fig:skewer}. Each transmission curve is calculated from
the sight-line in Figure \ref{fig:skewer} that bears the same number.
\label{fig:9spectra}}
\end{figure}

The absorption profile of a GRB afterglow probes a line of sight
through the IGM.  Most of the absorption from the IGM occurs at $<70
\; \Mpc$ from the host galaxy, and the neutral regions in the IGM that
lie closer to the GRB contribute more absorption than those further
away.  It is clear from Figure \ref{fig:slice} that different
sight-lines encounter vastly different spatial distributions of $x_H$.
To further illustrate this point, Figure \ref{fig:skewer} displays
nine randomly selected $70 \; \Mpc$ skewers originating at a potential
GRB site in the simulation volume.  These skewers are from the $z =
7.3$ snapshot of the simulation using model (i) for which the
volume-averaged neutral fraction is $\bar{x}_{H} = 0.5$ and the
mass-averaged neutral fraction is $0.4$.\footnote{The
ionisation fronts are much narrower than our grid size for
reionisation by POPII-like stars (as is assumed in these simulations).
Resolution effects smear out the simulation ionisation field, making
it deviate from a field of $0$s and $1$s.  This smearing does not
appreciably affect the results of the calculations reported here.}
Henceforth, $\bar{x}_{H}$ refers to the volume-averaged neutral
fraction.  To assign locations for GRBs in the simulation volume, we
assume that the $\dot{N}_{\rm ion}$ of a halo is proportional to the
GRB rate (which is true if the GRB rate traces the massive SFR and
$f_{\rm esc}$ is independent of halo mass).

Figure \ref{fig:9spectra} plots the IGM transmission redward of
Ly$\alpha$ for the nine GRB sight-lines shown in Figure
\ref{fig:skewer}.  To compute these curves, we use Equation (1) and
the density, ionisation, and velocity fields from the simulation.  The
observed transmission across these six lines of sight varies enormously, and
this variance needs to be accounted for in measurements of $\bar{x}_H$
using GRBs.  These calculations do not account for the evolution in
$\bar{x}_H$ along the line of sight owing to light-travel effects,
but instead compute the absorption from a skewer using a snapshot that
is fixed in time.  In the time for light to travel $50 \; \Mpc$, the
value of $\bar{x}_H$ in the simulations of model (i) changes by
$0.06$, $0.08$, and $0.13$ when $\bar{x}_H = 0.7$, $0.5$,
and $0.2$, respectively.  A proper treatment of the evolution
of $\bar{x}_H$ along a sight-line would hardly affect our
conclusions.

 For the nine sight-lines in Figure \ref{fig:skewer}, the best fit
$(\bar{x}_H, R_b)$ to the afterglow transmission redward of Ly$\alpha$
are respectively $(0.31, 6.6)$, $(0.35, 14)$, $(0.47, 2.3)$, $(0.51,
0)$, $(0.28, 0)$, $(0.21, 10)$, $(0.32, 22)$, $(0.45, 10)$, and
$(0.43, 2.5)$.\footnote{Note that the two skewers which are best fit
with $R_b = 0$ are anomalous, and the total fraction of lines of sight
that are best fit with $R_b = 0$ is a much smaller fraction than
suggested by this sample of nine sight-lines.}  While these two
parameters do not fully characterise the distribution of $\bar{x}_H$,
the actual value for $R_b$ of the GRB host bubble corresponds roughly
to the size given by these fits and the fitted value for $\bar{x}_H$
is within $0.3$ of the global value.  In reality, a DLA
will also affect the damping wing profile, and its absorption must be
accounted for simultaneously.  We incorporate this extra complication
in the following section.

\begin{figure}
 {\epsfig{file=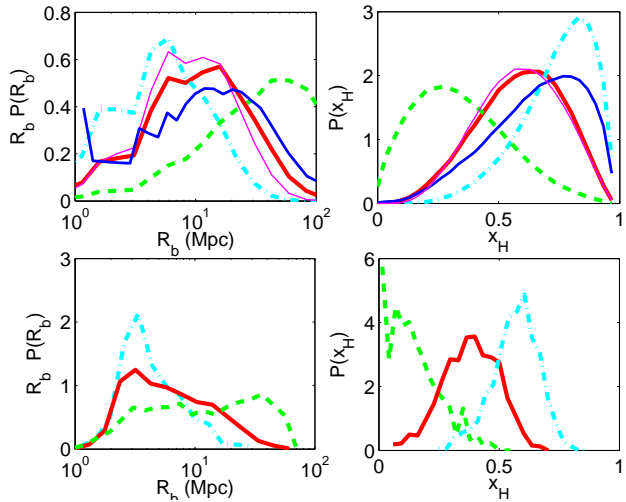, width=8.6cm}}
 \caption{Dashed, solid, and dot-dashed curves represent $\bar{x}_{H}
 = 0.2$, $0.5$ and $0.7$, respectively.  The thick curves are
 calculated using model (i), and, in the top panels, the medium width
 and thin solid curves (shown only for $\bar{x}_{H} = 0.5$) use model
 (ii) and model (iii). {\it Top Left Panel:} The PDF of $R_b$
 constructed from $10^6$ sight-lines, in which $R_b$ is defined as the
 length of a skewer originating at a GRB that passes through three
 grid cells with $x_H > 0.01$.  {\it Top Right Panel:} The PDF of
 $\bar{x}_H$ measured directly from $10^6$ lines of sight as discussed
 in the text. {\it Bottom Panels:} The same as the top panels except
 that the PDFs of $x_H$ and $R_b$ are constructed from fitting the
 afterglow transmission curves computed from $1000$ different
 simulation sight-lines with these parameters. }
 \label{fig:rb}\label{fig:xH}
 \end{figure}

The top left panel in Figure \ref{fig:rb} plots the probability
distribution of GRB host HII bubble radii, $P(R_b)$.  The GRB host
bubble radius is defined as the distance it requires for a skewer to
cross three simulation grid cells that are more than $1\%$ neutral.
The dashed, solid, and dot-dashed curves are $R_b \, P(R_b)$ for
$\bar{x}_{H} = 0.2$, $0.5$ and $0.7$, respectively.  The thick curves
are for model (i), and the medium width and thin solid curves (shown
for only $\bar{x}_{H} = 0.5$) adopt model (ii) and model (iii),
respectively.  These curves demonstrate that the function $P(R_b)$
depends more strongly on $\bar{x}_H$ than on the reionisation model.
When $\bar{x}_{H} \approx 0.5$, the typical GRB resides in a bubble of
size $R_b \approx 10 \; \Mpc$, and when $\bar{x}_{H} \approx 0.2$ this
number increases to $R_b \approx 60 \; \Mpc$.

  The top right panel in Figure \ref{fig:xH} shows the distribution of
 $\bar{x}_H$ along the line of sight to a GRB at fixed global
 $\bar{x}_H$.  To create this PDF, we generate $10^6$ skewers that
 originate at GRBs in the simulation volume and then tabulate
 $\bar{x}_H$ along each skewer, weighting points as $1/r^2$ where $r$
 is the distance from the source -- roughly the same weighting that is
 relevant to the damping wing optical depth at observed frequency
 $\nu_{\alpha}/(1 +z_g)$ (eqn. \ref{eqn:tau2}).  We do not include
 points inside the host bubble in this tabulation.

The wide breadth of the $\bar{x}_H$ distributions indicate that a
 single GRB cannot be used to directly measure the global value of
 $\bar{x}_H$ to a precision of $\delta \bar{x}_H \approx 0.3$.
 Despite the scatter in $\bar{x}_H$, a single GRB can be used to
 detect whether the IGM is neutral.  We focus on the probability of a
 GRB allowing the detection of reionisation in the next section.

 It is possible that GRBs occur in only the most massive galaxies
 (rather than in a typical one, as we have assumed thus far).  If only
 halos with $m > 10^{10} \; \Msun$ are able to produce GRBs with a
 rate proportional to the halo mass, again weighting these halos by
 their SFR, we find that the average bubble size increases by a
 comparable amount to the size increase between model (i) and model
 (ii) shown in Figure \ref{fig:rb}.  However, we find that the low-end
 tail of $P(R_b)$ starts to disappear with increasing $m$, even though
 the peak of the PDF does not shift significantly.


 Thus far, we have discussed the distribution of $\bar{x}_H$ and $R_b$
 measured directly from the simulations.  For a GRB afterglow,
 $\bar{x}_H$ and $R_b$ are measured by fitting these parameters to the
 afterglow damping wing (even though this parameterisation for the
 distribution of $x_H$ along a sight-line is highly simplified).  The
 bottom panels in Figure \ref{fig:xH} show the PDF of $R_b$ and
 $\bar{x}_H$ from fits to $1000$ transmission curves computed from
 GRBs in the simulation of model (i). We ignore the effect of DLA
 absorption in these fits.  These PDFs are notably different from
 those in the top panels -- favouring smaller $R_b$ and lower
 $\bar{x}_H$.  While this procedure returns biased values for $R_b$
 and $\bar{x}_H$ relative to the values we estimate in the top panel,
 we find that this two parameter model provides an excellent fit to
 the absorption profile (as we discuss in Section \ref{fits}).


 \section{Fitting the Damping Wing in GRBs}
 \label{fits}
 This section quantifies the circumstances under which a GRB afterglow
 can be used to detect a neutral IGM.  Previous sections have
 illustrated how the patchiness of the IGM and the strength of the
 host DLA complicate this measurement.  In addition, there are other
 uncertainties such as the power-law slope and amplitude of the
 intrinsic GRB spectrum, the precise redshift of the GRB and of the
 DLA, and dust absorption and metal line contamination from the host
 galaxy or from intervening systems.  A systematic discussion of these
 issues is presented in \citet{totani06}.  To rule out an ionised IGM
 with a GRB afterglow, a multi-parameter fit to its spectrum that
 assumes an ionised IGM and that accounts for the aforementioned
 uncertainties must provide a poor fit.

\subsection{Fits to a Toy Absorption Model}

\begin{figure}
 {\epsfig{file=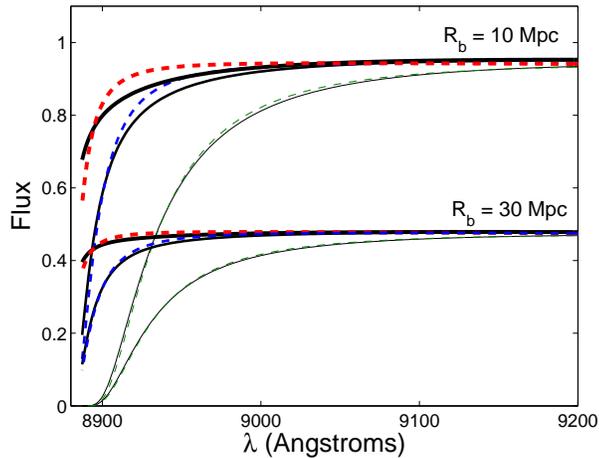, width=8.7cm}}
 \caption{Afterglow spectrum of a hypothetical $z = 6.3$ burst,
 computed with $\bar{x}_H = 0.5$ and $N_{\rm HI} = 10^{19}$ cm$^{-2}$
 (thick solid curves), $N_{\rm HI} = 10^{20}$ cm$^{-2}$ (medium-width
 solid curves), and $N_{\rm HI} = 10^{21}$ cm$^{-2}$ (thin solid
 curves) for the specified $R_b$.  The dashed curves are the best fits
 for a model that includes only DLA absorption, fitted to $\Delta
 \lambda = 200\; (1+z_g) = 1500 \AA$.  The normalisation of the two
 sets of curves is adjusted to aid viewing. \label{fig:fits}}
\end{figure}

 A toy example of such a multi-parameter fit that accounts for some of
 these uncertainties is shown in Figure \ref{fig:fits}.  In this
 figure, the simple case of a bubble with size $R_b$ in a homogeneous
 IGM with neutral fraction $\bar{x}_H$ is shown.  The solid curves are
 absorption models that have the intrinsic parameters $\beta = 1.25$,
 $z_g= z_{\rm DLA} = 6.3$,\footnote{The assumption that $z_g= z_{\rm
 DLA}$ is probably reasonable for GRBs in high-redshift galaxies,
 where the circular velocity of these halos is typically $\sim 25$ km
 s$^{-1}$ \citep{barkana04b}.} $\bar{x}_H = 0.5$ and $N_{\rm HI} =
 10^{19}$ cm$^{-2}$ (thick curves), $N_{\rm HI} = 10^{20}$ cm$^{-2}$
 (medium-width curves), or $N_{\rm HI} = 10^{21}$ cm$^{-2}$ (thin
 curves).\footnote{We include the prior $0.75 < \beta < 1.75$ in the
 fits: $\beta$ will likely be constrained by earlier photometric
 observations redward of the absorption, as was the case for GRB050904
 \citep{totani06}. In addition, we exclude wavelengths within $400
 ~{\rm km s}^{-1}$ of the host galaxy because such regions could be
 affected by Ly$\alpha$ emission from the galaxy as well as resonant
 absorption by infalling material.}  We then fit to these model
 curves. Our fitting weights all frequency bins equally.  The presence
 of night-sky OH lines and the wavelength dependence of the CCD
 sensitivity should make the weighting slightly non-uniform in a real
 observation.

 The dashed curves in Figure \ref{fig:fits} are fits over $\Delta
 \lambda = 200\; (1+z_g)$ to a model that only include DLA absorption.
 This model fits for the parameters $N_{\rm HI}$, $\beta$, and $A$.
 The absorption owing to a substantially neutral IGM is more easily
 fit with \emph{just} a DLA model as $R_b$ increases or as $N_{\rm
 HI}$ increases.  If we had fit over a larger $\Delta \lambda$, the
 dashed and solid curves would differ by less at wavelengths where the
 absorption is important.  The fits in Figure \ref{fig:fits} assume
 that $z_{\rm DLA}$ is known from metal lines, as was the case for
 GRB050904.  For GRBs in more metal-poor galaxies or for GRBs with
 smaller DLAs, this may not always be possible.  When fits include
 $z_{\rm DLA}$ as a free parameter, a neutral IGM is much harder to
 distinguish from an ionised one \citep{barkana04b}.

 To quantify the capability of distinguishing different absorption
 models, we define the measure
 \begin{equation}
   \langle \Delta \chi^2 \rangle= \sum_{i = 1}^{N} \; \frac{\left[ F_2(\lambda_i) - F_1(\lambda_i) \right]^2}{\sigma(\lambda_i)^2},
 \end{equation}
 where $F_1$ and $F_2$ are the afterglow fluxes for models $1$ and
 $2$.  This quantity represents the average difference in $\chi^2$
 between two models for a spectrum that has noise $\sigma(\lambda_i)$
 in channel $\lambda_i$ and that has $N$ wavelength channels, assuming
 the data is drawn from model $1$ or $2$.

 When comparing a model with a neutral IGM to an ionised one, two
 models can be distinguished at $X$ confidence level (C.L.)  if $X$ of
 the total likelihood is contained between the maximum likelihood in
 the neutral IGM model and $\exp[-\langle \Delta \chi^2 \rangle/2]$ of
 that likelihood value (assuming Gaussianity). For a $4$ parameter
 fit, two models that differ by $\langle \Delta \chi^2 \rangle = 8$
 signifies that model 1 is, in the mean, preferred at $91\%$ C.L. over
 model 2 ($\langle \Delta \chi^2 \rangle = 5$, $70\%$ C.L. and
 $\langle \Delta \chi^2 \rangle = 12$, $98\%$ C.L.).  The significance
 levels at fixed $\Delta \chi^2$ for fits with three and five
 parameters are similar.  In what follows, we quote $\langle \Delta
 \chi^2 \rangle$ in terms of $\Delta \lambda_i = 3 \; \AA$ and
 $\sigma_{\cal F} = 0.1$, where $\Delta \lambda_i$ is the width of a
 frequency channel and $\sigma_{\cal F}$ is the standard deviation in
 the flux in each spectral channel in units of $A$.  These are
 approximately the values of $\Delta \lambda_i$ and $\sigma_{\cal F}$
 for the Subaru FOCAS spectroscopic observation of GRB050904.

 First, we consider the case in which the absorption spectrum is
 parameterised by $N_{\rm HI} \approx 10^{19}$ cm$^{-2}$, $\bar{x}_H =
 0.5$, and $R_b = 10$ Mpc (similar to the fits in
 Fig. \ref{fig:fits}).$^8$  We fit to $\Delta \lambda = 50 \;(1+z_g) \;
 \AA$ redward of source-frame Ly$\alpha$.  This fit results in
 $\langle \Delta \chi^2 \rangle = 5 \times (3 \AA / \Delta \lambda_i
 )\times (0.1/\sigma_{\cal F})^2$ between the model for a neutral IGM
 and the model for an ionised IGM.  Next, we fit to a model with the
 same specifications as the previous fit but with $N_{\rm HI} \approx
 10^{20}$ cm$^{-2}$ [$N_{\rm HI} \approx 10^{21}$ cm$^{-2}$].  This
 fit results in $\langle \Delta \chi^2 \rangle = 3 \; [0.2] \times (3
 \AA/\Delta \lambda_i) \times (0.1/\sigma_{\cal F})^2$.  Therefore,
 $\Delta \lambda_i \lesssim 3 \; \AA$ and $\sigma_{\cal F} \lesssim
 0.1$ are required for any hope of detecting a neutral IGM for the
 $N_{\rm HI} \approx 10^{19}$ cm$^{-2}$, $x_H \approx 0.5$, and $R_b =
 10$ Mpc case, and even better sensitivity is required for the other
 two examples.

Let us investigate how the quoted $\langle \Delta \chi^2 \rangle$ depend on the
assumptions we have made.  For this discussion, we again assume an
absorption model with $\bar{x}_H = 0.5$, $R_b = 10$ Mpc, and $N_{\rm
HI} \approx 10^{20}$ cm$^{-2}$, and we assume a measurement with
$\Delta \lambda_i \approx 3 \; \AA$ and $\sigma_{\cal F} \approx 0.1$.
\begin{itemize}
\item If we restrict $\beta$ to $1.0 <\beta < 1.5$ (rather than to $0.75
<\beta < 1.75$, as before), $\langle \Delta \chi^2 \rangle$ increases
from $3$ to $4$.  If we fix $\beta = 1.25$, then $\langle \Delta
\chi^2 \rangle = 5$.  

\item If we allow $z_{\rm DLA}$ to vary
and fix $\beta = 1.25$, $\langle \Delta \chi^2 \rangle$ decreases to $1$.  
\item If we instead fit the first $ \Delta \lambda =200 \; (1+z_g) \;
\AA$ redward of Ly$\alpha$ rather than $ \Delta \lambda = 50 \;(1+z_g) \;
\AA$, $\langle \Delta \chi^2 \rangle$ increases from $3$ to $6$.
\end{itemize}
The increase in $\langle \Delta \chi^2 \rangle$ with bandwidth owes to
tighter constraints on $A$ and $\beta$, the parameters that determine
the spectrum far redward from the Ly$\alpha$ absorption feature.
These tighter constraints breaks degeneracies between these parameters
and the parameters that determine the absorption.  The appropriate
choice of bandwidth will not always be clear-cut; fits should not
include wavelengths where the dust absorption, the instrumental
response, and other factors are not understood to better than the
difference between the two absorption models.  Already for $\Delta
\lambda \approx 300\; \AA$, such calibration is necessary to the $1\%$
level (Fig. \ref{fig:fits}).  In addition, even uncertainties in the
cosmological parameters make a difference in distinguishing models at
the percent level \citep{miralda98}.

\subsection{Fits to the Simulated Absorption Spectra}
 Thus far, we have considered fits to a toy model for IGM
 absorption. To realistically model IGM absorption we calculate mock
 spectra from simulation skewers of reionisation model (i) and we set
 $\beta = 1.25$.  We fit this mock data with different absorption
 models in what follows.

Despite the complicated distribution of
 neutral hydrogen along sightlines, a simple model with just the
 parameters $R_b$ and $\bar{x}_H$ (in addition to parameters that
 account for DLA absorption and the intrinsic afterglow spectrum)
 typically provides an excellent fit to the mock data. This is surprising
 because of the highly simplified nature of this model.  The $\langle
 \Delta \chi^2 \rangle$ between the true model -- the absorption
 profile calculated from the simulation -- and the best fit
 neutral IGM model, which fits for $R_b$ and $\bar{x}_H$, typically
 range between $10^{-3} - 10^{-7} \times (3 \AA /\Delta \lambda_i)
 \times (0.1/\sigma_{\cal F})^2$ for $N_{\rm HI} = 10^{20}\; {\rm
 cm}^{-2}$.  These small values mean that $\sigma_{\cal F} \approx
 10^{-2}- 10^{-4}$ is required to rule out this $2$-parameter IGM
 absorption model (measurements $\approx 10-1000$ times more sensitive
 than those for GRB050904).

 \begin{figure}
 \epsfig{file=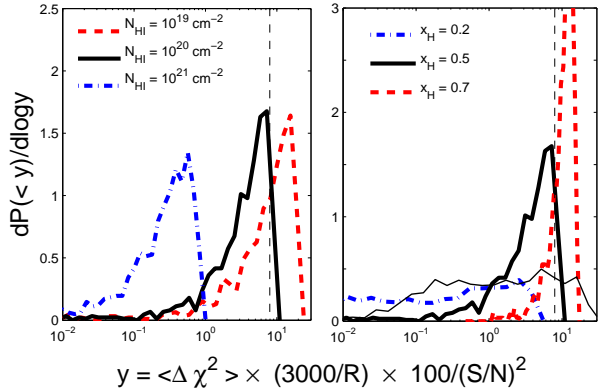, width=8.4cm}
 \caption{PDFs of $\langle \Delta \chi^2 \rangle$, where $\langle
 \Delta \chi^2 \rangle$ is the average difference in $\chi^2$ between
 the ``true'' model (which fits $\bar{x}_H$ and $R_b$) and a model
 with only DLA absorption.  These PDFs are generated by fitting $1000$
 sight-lines from the simulation of model (i) at the specified $\bar{x}_H$.
 Note that the parameter $y$ that is plotted on the abscissa is scaled
 to be independent of $S/N$ and $R$.  The left panel shows the
 dependence of the PDF on $N_{\rm HI}$, using spectra from a snapshot
 with $\bar{x}_H = 0.5$, and the thick curves in the right panel show
 the dependence on $\bar{x}_H$, assuming $N_{\rm HI} = 10^{20}\; {\rm
 cm}^{-2}$.  The thin black curve in the right panel folds in the
 observed distribution of $N_{\rm HI}$ and uses spectra with
 $\bar{x}_H= 0.5$.  A neutral IGM is detected on average at $>84\%$
 confidence level when $\langle \Delta \chi^2 \rangle > 8$ (marked by
 the vertical dashed lines for $R = 3000$ and $S/N = 10$). 
 \label{fig:chisq}}
 \end{figure}

 Figure \ref{fig:chisq} shows the distribution of $\langle \Delta
 \chi^2 \rangle$ between fits to the simulated spectra with the
 parameters $R_b$, $\bar{x}_H$, $N_{\rm HI}$, $A$ and $\beta$ versus
 fits with just the last three parameters (i.e., a DLA and an ionised
 IGM).  These fits are to a wavelength interval that spans $50\;(1+z)
 \;\AA$ redward of rest-frame Ly$\alpha$. If the fits with $\bar{x}_H
 = 0$ are disfavoured by the data (i.e., if $\langle \Delta \chi^2
 \rangle \gtrsim 8$), then a neutral IGM will be favoured by the data.
 The parameter $y$ that is plotted on the abscissa in Figure
 \ref{fig:chisq} is scaled to be independent of $S/N$ (we define $S/N
 \equiv \sigma_F^{-1}$) and spectrograph resolution $R$ ($R \equiv
 \lambda_i/\Delta \lambda_i$).  For an observation like the FOCAS
 observation of GRB050904 ($R = 3000$ and $S/N = 10$), a DLA with
 $N_{\rm HI} \lesssim 10^{20}$ cm$^{-2}$ is needed to have a good
 chance of detecting a neutral IGM when $\bar{x}_H \approx 0.5$ (left
 panel in Fig. \ref{fig:chisq}).  For $\bar{x}_H < 0.5$, a higher
 signal-to-noise measurement is required.

The results in Figure \ref{fig:chisq} depend on the parameters we fit
as well as the wavelength range included in the fit.  If we fix the
parameter $\beta = 1.25$ instead of fitting for it, the PDFs in this
figure are, for the most part, unchanged.  If we fit a larger range in
wavelength of $\Delta \lambda = 200\;(1+z) \;\AA$, the fits improve
slightly -- shifting the histogram by factor of $\approx 1.5$.

 The thin solid curve in the right panel of Figure \ref{fig:chisq} is
 the PDF of $\langle \Delta \chi^2 \rangle$ for spectra computed for
 $\bar{x}_H = 0.5$ and where we have folded in the distribution of DLA
 $N_{\rm HI}$ values observed in lower redshift GRB-host systems.  We
 assume that the cumulative distribution of DLAs is a power-law of the
 form $N_{\rm HI}^{0.3}$ for $N_{\rm HI} < 10^{21.5}$ cm$^{-2}$, which
 is consistent with observations \citep{chen07}.  Half of GRB DLAs
 have $N_{\rm HI} > 10^{21.5}$ cm$^{-2}$, but we do not include these
 bursts in this computation.  These bursts will contribute to the
 small $\langle \Delta \chi^2 \rangle$ tail of the PDF.

Let us assume there exists an observation with similar sensitivity to
 the Subaru FOCAS spectrum of GRB050904, that the distribution of DLAs
 is the same as found at lower redshift, that the redshift of the GRB
 is known via metal lines, and that the GRB originates from a redshift
 at which $\bar{x}_H \approx 0.5$.  The spectrum of this GRB can be
 used to detect a partly neutral IGM at $97\%$ C.L. ($\langle \Delta
 \chi^2 \rangle > 12$) approximately $5\%$ of the time (and, for an
 observation with $3$ times the sensitivity, $\approx 25\%$ of the
 time) if the fit uses $\Delta \lambda =
 50\;(1+z)\;\AA$.\footnote{These numbers assume that bursts with
 $N_{\rm HI} > 10^{21.5}$ cm$^{-2}$ do not provide a positive
 detection of a neutral IGM.}  If the fit uses $\Delta \lambda =
 200\;(1+z)\;\AA$, this becomes $\approx 10\%$ of the time (and, for
 an observation with $3$ times the sensitivity, $\approx 30 \%$ of the
 time).

An earlier spectroscopic pointing could result in higher $S/N$ values
 than the fiducial value of $10$ because the optical afterglow fades
 as a power-law in time with slope $\approx -1.2$ \citep{liang06}.
 The same Subaru observation taken $\approx 8$ hr after GRB050904
 would have resulted in $10\times$ the sensitivity and a detection of
 reionisation the majority of the time.  In 2013, JWST is projected to
 be in orbit.  Let us assume a $z = 8$ GRB is observed with a $10^4$
 s integration using the JWST spectrograph NIRSpec operating in its highest
 resolution mode of $R = 2700$.  To detect the GRB with $S/N = 10$ in
 each frequency channel requires a minimum flux of $F(\lambda) =
 2\times10^{-19}$ erg cm$^{-2}$ s$^{-1}$
 $\AA^{-1}$.\footnote{http://www.stsci.edu/jwst/instruments/nirspec/}
 This minimum flux is approximately five times the sensitivity of the
 Subaru FOCAS observation of GRB050904.  However, while ground based
 observations can potentially target GRB afterglows minutes after
 their trigger, JWST will only be able to slew to high-redshift
 afterglows $\sim 1$ day after the burst \citep{gardner06}.

 Even though the calculations in this section were for $z \approx
 6-8$, similar conclusions hold for higher redshifts reionisation
 scenarios.  The amount of absorption does increase with increasing
 $z_g$ because the damping wing optical depth scales as $\approx (1
 +z_g)^{3/2}$ (eqn. \ref{eqn:tau2}).  However, the morphology of
 reionisation depends weakly on redshift for stellar reionisation
 scenarios -- bubbles have a similar comoving size distribution at a
 fixed $\bar{x}_H$ independent of when reionisation happens
 \citep{mcquinn06b}.  For reasonable redshifts over which reionisation
 could occur, the amount of damping wing absorption predicted by our
 simulations is similar and our conclusions do not change.

 As a final point, the distribution of the best-fit $R_b$ and
 $\bar{x}_H$ from the fits that include DLA absorption are very
 similar to the distribution in Figure 5 in which the effect of a DLA
 was ignored.  Therefore, the presence of a DLA does not significantly
 bias the best fit values for $R_b$ and $\bar{x}_H$. 

 \section{GRB050904}
 \label{GRB050904}

\begin{figure}
{\epsfig{file=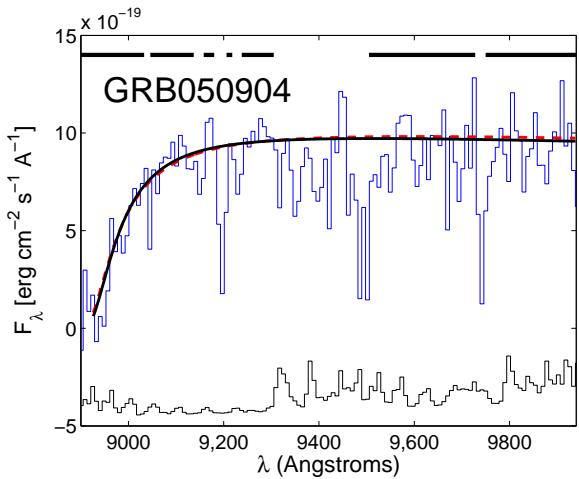, width=8.5cm}}
\caption{Binned afterglow spectrum of GRB050904 (each $2.67\; \AA$
wavelength channel has been binned into $8\; \AA$ pixels), as well as
the best fit models for $x_H = 0$ (solid curve) and $x_H= 1$ (dashed
curve).  The bottom curves are the 1-$\sigma$ errors on each binned
pixel with an offset of $-5\times 10^{-19}$, and the thick horizontal
lines at the top of the figure indicate the wavelengths that were
included in the fits.  Both fits require a DLA with column density
$N_{\rm HI} \approx 10^{21.5} \; {\rm cm}^{-2}$.}
\label{fig:data}
\end{figure}

The $z = 6.3$ GRB, GRB050904, is the burst with the highest identified
redshift \citep{kawai05, haislip06, totani06}. A $4$ hour observation
of this burst was taken using the FOCAS spectrograph on Subaru $3.4$
days after the prompt gamma ray emission.  The spectrum from this
observation is shown in Figure \ref{fig:data}.  \citet{totani06}
showed that this spectrum is well-fitted with a DLA of column density
$N_{\rm HI} \approx 10^{21.6}$ cm$^{-2}$ and that this GRB disfavours
additional absorption owing to a neutral IGM, deriving the constraint
$\bar{x}_H < 0.6$ at $95\%$ C.L. This section investigates the
constraints on $\bar{x}_H$ from GRB050904 if a patchy reionisation
process consistent with simulations is taken into account.

First, we have re-performed the analysis of \citet{totani06},
accounting for galactic absorption of $E(B-V) = 0.06$ and masking the
same regions of the spectrum as in \citet{totani06} because of metal
line contamination.  (The thick horizontal lines at the top of Figure
\ref{fig:data} indicate the wavelength ranges that were fit.)  The
solid absorption curve in Figure \ref{fig:data} is the best fit for an
ionised IGM ($\bar{x}_H= 0$) with a DLA and the dashed curve is the
best fit with a DLA plus a \emph{fully} neutral IGM.  Both fits to the
spectrum employ the same estimated errors as in \citet{totani06} and
minimise $\chi^2$ with two parameters, $N_{\rm HI}$ and $A$, holding
all other parameters fixed.  In particular, we fix $\beta = 1.25$ as
was done in \citet{totani06}.  This value for $\beta$ is consistent
with $\beta = 1.25 \pm 0.25$, which was derived in
\citet{tagliaferri05} and is consistent with the value derived in
\citet{kann07} from all publicly available data.\footnote{The
\citet{tagliaferri05} value for $\beta$ was derived from photometric
observations with the Very Large Telescope taken $1.2$ days after the
prompt emission. \citet{totani06} argued that the softness of the
$\beta$ derived in \citet{tagliaferri05} indicates that $\beta$ was
determined by the power-law index of the electrons and, therefore,
should not evolve significantly between the time of the measurement
from \citet{tagliaferri05} and their measurement (i.e., other breaks
in the synchrotron spectrum are at longer wavelengths).}  The value of
$\Delta \chi^2$ is not significantly changed if we instead fit with
$\beta$ adopting the prior $1.0 <\beta < 1.5$ (the best fit for both
models then prefers $\beta = 1$), and $\Delta \chi^2$ shrinks
significantly if no prior is placed on $\beta$.\footnote{If we ignore
photometric constraints on $\beta$ and just fit it as a free
parameter, the ionised IGM model prefers $\beta = -0.6$ at $98\%$
C.L. ($\Delta \chi^2 = 10$) over a model with $\beta = 1.25$.  While
it is hard to imagine $\beta$ evolving to this small value from $\beta
= 1.25$, this nevertheless may indicate that there is a problem with
the model for the data.  Fits to the spectrum that do not place a
prior on $\beta$ do not prefer ionised IGM models over neutral ones.}

 The $\chi^2$ values for the two models displayed in Figure
\ref{fig:data} are $277.3$ for the ionised model and $287.0$ for the
neutral case.\footnote{To emulate the results of \citet{totani06}, the
error bars have been renormalised by the factor of $0.848$ from their
originally estimated value such that the $\chi^2/{\rm d.o.f.}  \approx
1$ for the best fit ionised model, where the number of degrees of
freedom (d.o.f.) equals $268 - 3$.  \citet{totani06} uses this
rescaling to correct for uncertainty in the errors. The residuals for
the best fit ionised model are close to a Gaussian with the correct
width after this procedure (Fig. $3$ in \citealt{totani06}).}  Even
though these two $3$-parameter models are difficult to distinguish
visually, it is the case that $\Delta \chi^2 =9.7$, which at face
value implies that the ionised model is preferred over the neutral
model at $98\%$ C.L. \citep{totani06}.  Interestingly, much of the
contribution to $\Delta \chi^2$ between the neutral and ionised models
comes from wavelengths that are not significantly affected by IGM and
DLA absorption.  If we tabulate $\Delta \chi^2$ for the best fit
models in Figure \ref{fig:data}, only including the wavelengths within
$200 \; \AA$ of Ly$\alpha$ -- the wavelengths most affected by
absorption -- in the summation, the two model fits differ by $ \Delta
\chi^2 = 4$.  If we include $700\; \AA$ redward of Ly$\alpha$, the two
model fits in Figure \ref{fig:data} differ by $ \Delta \chi^2 = 5$.
The significant contribution to $\Delta \chi^2$ from redward of the
Ly$\alpha$ absorption stems from the fact that the neutral IGM model
prefers a $2.4\%$ higher normalization to compensate for the
additional effect of IGM absorptions. As a consequence, the unabsorbed
part of the spectrum needs to be modelled to a relative precision of a
percent over a wavelength range with $\Delta \lambda / \lambda \approx
0.1$ for a correct interpretation of the data.

To distinguish models that differ at the $2\%$ level over hundreds of
angstroms posits that intergalactic metal and sky lines have been
properly masked, that the dust reddening of the Milky Way and the host
galaxy has been accurately accounted for, that the relative
calibration of and errors on the measured flux are accurate, and that
the model for the intrinsic spectrum is correct. \citet{totani06}
investigated many of these uncertainties and argued that the FOCAS
spectrum of GRB050904 can be used to distinguish the models in Figure
\ref{fig:data}.\footnote{\citet{totani06} also used the transmission
in the Ly$\beta$ forest as an additional handle to discriminate
between the best fit for a model that assumed DLA absorption and an
ionised IGM and the best fit model that assumed the absorption owed
\emph{entirely} to a neutral IGM.  The later fit required the redshift
of the host galaxy to be significantly redward of the best fit
redshift of the DLA absorption lines, which was inconsistent with the
transmission in the Ly$\beta$ forest.  This example is a rather
exceptional case, and for most bursts Ly$\beta$ will not provide an
additional handle.  In addition, patchy reionisation allows for
transmission blueward of GRB-frame Ly$\beta$.  Therefore, the
\citet{totani06} argument can be applied only when the model value for
$\bar{x}_H$ is larger than the fraction of pixels that have
transmission.}

The above analysis did not fit for $R_b$ and $\bar{x}_H$, which must
be done to place constraints on $\bar{x}_H$ for GRBs during the Epoch
of Reionisation.  However, while GRB050904 may favour an ionised
universe over a neutral universe, it is more difficult to place a
constraint on intermediate $\bar{x}_H$ with GRB050904, especially in
light of the large bubble sizes during reionisation.  If we fit a
model that fixes $\bar{x}_H = 0.5$ (about as low an $\bar{x}_H$ as is
plausible at $z = 6.3$ \citep{lidz07}) and $R_b = 10~\Mpc$ [$30~\Mpc$]
-- again fitting with $A$ and $N_{\rm HI}$ and fixing $\beta$ --, then
the $\Delta \chi^2$ between a model with an ionised universe and this
model is $3.1$ [$2.5$], which is not statistically significant.  To
place any constraint on $\bar{x}_H$ for GRB050904, one must account
for the covariance of $R_b$ and $\bar{x}_H$ (i.e., lower $\bar{x}_H$
have smaller $R_b$).  We have not taken the additional step of
accounting for this covariance in our analysis because it will not
lead to interesting constraints on $\bar{x}_H$ from GRB050904.  For
future bursts that prefer no IGM absorption, such an anlysis will be
essential.  In any case, accounting properly for patchy reionisation
would significantly weaken the $95\%$ C.L. constraint $\bar{x}_H <
0.6$ at $z =6.3$ that \citet{totani06} derives from GRB050904.

\section{The Damping Wing in QSOs and Galaxies}
\label{qsoandgalaxies}
GRBs are not the only beacons in which the signature of a neutral IGM
can be observed in their continuum emission.  Rather than wait for a
high-redshift GRB to occur, hundreds of $z >6$ galaxies and QSOs have
already been found.  Perhaps these objects can be used to detect a
neutral IGM.  However, since the observed population of QSOs and
galaxies are more biased tracers of the high-redshift Universe than
are GRBs (assuming that GRBs trace star formation), they sit in larger
bubbles such that the damping wing absorption is smaller, on average.
The effective size of an HII region for one of the known $z \approx 6$
quasars, assuming that the IGM is significantly neutral at this
redshift, is predicted to be $\gtrsim 50 ~\Mpc$.  This number accounts for
these rare objects being in the most overdense regions in which
reionisation occurs earlier \citep{lidz07}.  Such large HII regions
make searches that target the red damping wing in the highest redshift
quasars hopeless.\footnote{Rather than analyse wavelengths redward of
the QSO Ly$\alpha$ line, damping wing absorption from neutral patches
in the IGM would also affect the Ly$\alpha$ forest of these QSOs
\citep{mesinger07}.  Owing to little forest transmission at $z >6$,
this feature would be difficult to detect \citep{lidz07}.}

Galaxies will sit in smaller HII regions than QSOs. The average bubble
size around galaxies is larger than those for GRBs.  As noted in \S
\ref{reionisation}, the difference in the typical bubble size is
typically less than a factor of two between halos of $m \approx
10^{10-11} \; \Msun$ (roughly the mass of halos of spectroscopically
confirmed $z >6$ galaxies) and GRB-hosting halos.

However, there are two significant complications.  First, because galaxies are much fainter, the
continuum emission for high-redshift galaxies is difficult to detect
spectroscopically with existing facilities.  Stacking galaxy spectra
is necessary to have any hope of detecting a damping wing feature with
current data.  Another significant complication is that the continuum
of galaxies is far from a simple power-law.  Such uncertainties must be
accounted for in any analysis that attempts to obtain $\bar{x}_H$ from
the damping wing absorption profile of a galaxy.


\section{Conclusions}
GRBs are the most luminous sources at high redshifts, and their smooth
power-law spectra are ideal for isolating the effects of absorption
owing to a neutral IGM.  However, observations must
separate the impact of IGM absorption from that of a DLA to detect a
neutral IGM. If no damping wing feature from IGM absorption is
detected in the spectrum of a high-redshift GRB, one must be careful
to conclude that reionisation is complete at the redshift of interest.

We have shown that there is a wide probability distribution of HII
region sizes, and that there is large variation in the distribution of
$\bar{x}_H$ along different sight-lines.  A non-detection of neutral
hydrogen from a GRB afterglow spectrum might arise because the GRB
host galaxy sits within a large HII region.  If absorption owing to a
neutral IGM is detected, it will be impossible to infer $\bar{x}_H$
from a single GRB to better than $\delta \bar{x}_H \sim 0.3$ because
of the patchiness of reionisation.

Assuming an observation with similar sensitivity to the Subaru FOCAS
spectrum of GRB050904, that the distribution of DLAs is the same as
found at lower redshifts, and that the redshift of the GRB is known, a
GRB from a redshift at which $\bar{x}_H \approx 0.5$ can be used to
detect a partly neutral IGM at $98\%$ C.L.  $\approx 10\%$ of the time
(and, for an observation with $3$ times the sensitivity, $\approx
30\%$ of the time).  If $\bar{x}_H < 0.5$, these percentiles for
detection are even smaller.  Weaker DLAs enhance the probability of
detecting a neutral IGM, but too weak of a DLA may prevent a precise
redshift determination, which is essential for tight constraints on
$\bar{x}_H$.

Since the $z = 6.3$ burst GRB050904 has a DLA with $N_{\rm HI} =
10^{21.6}$ cm$^{-2}$, the absorption on the red side of the line is
dominated by the DLA \citep{totani06}.  While this burst may favour a
model with an ionised universe over a neutral universe
\citep{totani06}, a weaker DLA is necessary to be able to constrain
$\bar{x}_H$ in the spectrum of a high-redshift GRB.  This is
particularly true if the GRB occurs within a large HII region.

GRB050904 was observed spectroscopically by the Subaru telescope $3.4$
days after the prompt gamma ray emission.  If this afterglow had been
observed hours after the burst, the flux would have been more than an
order of magnitude larger.  To detect a neutral IGM, it is crucial for
optical and near-infrared spectrographs to observe candidate
high-redshift GRBs as soon as possible after the prompt gamma ray
emission.  A high signal-to-noise ratio is critical to
distinguish IGM absorption from that arising from a DLA.  Such an
observing programme is worthwhile given the promise that GRBs have as
probes of the epoch of reionisation.


\section{acknowledgments}
We would like to thank T. Totani for providing the Subaru afterglow
data for GRB050904 as well as for his comments on the manuscript and
for answering numerous questions about the measurement.  We also thank
Mark Dijkstra, Claude-Andr{\'e} Faucher-Gigu{\`e}re, Alexandre
Tchekhovskoi, and Hy Trac for many interesting discussions and the
anonymous referee for useful comments on the manuscript.  MM
acknowledges support through an NSF graduate student fellowship.  The
authors are also supported by the David and Lucile Packard Foundation,
the Alfred P. Sloan Foundation, and grants AST-0506556 and NNG05GJ40G.

\begin{appendix}

\end{appendix}

\bibliographystyle{mn2e}

\end{document}